\documentclass[twocolumn,aps,showpacs,prb,tightenlines,amsmath,amssymb,superscriptaddress]{revtex4}
\usepackage{graphicx}
\usepackage{dcolumn}
\usepackage{bm}
\begin{document}
\title{Anisotropic spin transport in GaAs quantum wells
in the presence of competing Dresselhaus and Rashba
spin-orbit coupling}
\author{J. L. Cheng}
\affiliation{Hefei National Laboratory for Physical Sciences at
Microscale, University of Science and Technology of China, Hefei,
Anhui, 230026, China}
\affiliation{Department of Physics,
University of Science and Technology of China, Hefei, Anhui,
230026, China}
\altaffiliation{Mailing address.}
\author{M. W. Wu}
\thanks{Author to whom all correspondence should be addressed}
\email{mwwu@ustc.edu.cn.}
\affiliation{Hefei National Laboratory
for Physical Sciences at Microscale, University of Science and
Technology of China, Hefei, Anhui, 230026, China}
\affiliation{Department of Physics, University of Science and
Technology of China, Hefei, Anhui, 230026, China}
\author{I. C. da Cunha Lima}
\affiliation{Hefei National Laboratory for Physical Sciences at
Microscale, University of Science and Technology of China, Hefei,
Anhui, 230026, China}

\date{\today}

\begin{abstract}

Aiming at the optimization of the spin diffusion length in (001) GaAs
quantum wells, we explore the effect of the anisotropy of the
spin-orbit coupling on the competition between the Rashba and the
Dresselhaus spin-orbit couplings by solving the kinetic spin Bloch
equations with the electron-phonon and the electron-electron scattering
explicitly included. For identical strengths of the Rashba and the
Dresselhaus spin-orbit couplings, the spin diffusion length shows strong
anisotropy not only for the spin polarization direction but also for the
spin diffusion direction.  Two
special directions are used seeking for the large diffusion length:
($\overline{1}$10) and (110). Without the cubic term of
the Dresselhaus spin-orbit coupling
and with the identical Dresselhaus and Rashba
strengths, infinite diffusion lengths can be obtained {\em either} for the
spin diffusion/injection direction along $(\bar110)$, regardless of the
direction of spin polarization, {\em or} for the spin polarization
along $(110)$, regardless of the direction of the spin diffusion/injection.
 However, the cubic Dresselhaus term cannot be neglected,
resulting in a finite spin diffusion length which decreases with the
temperature and the electron density. The anisotropy for the spin
diffusion direction and spin polarization direction is maintained.
For the spin diffusion/injection direction along ($\bar{1}$10), the spin
diffusion length increases first with the increase of the Rashba
strength (from 0) which can be tuned by the external gate voltage; when the
Rashba strength is slightly smaller than (in stead of equal to)
 the Dresselhaus strength, the
diffusion length reaches its maximum, followed by a decrease with
further increase of the Rashba strength.
\end{abstract}

\maketitle

\section{Introduction}

The control of the electron spin  in nanoscale is an object of
great interest in the field of
spintronics.\cite{spintronics,zutic} The knowledge of how the spin
evolves as the electron goes through a nano-device is a key
element for the purpose of using the spin degree of freedom as a
mechanism  for information transfer and processing.
Randomization of the electron spin, however, occurs due to the
spin-orbit coupling as the electron wave vector changes during and
after scattering processes. Between scattering events, the precession
frequency of the spin about the local magnetic field also changes
from electron to electron depending on the wave vector. To
overcome these detrimental effects, many efforts have been made to
understand the spin diffusion and the spin
relaxation,\cite{wengwu1,wengwu2,wu1,wengwushi,wu,aver1,aver2,
John,Loss1,Loss2,winkler1,lechner,Cartoixa,Ehud} in particular after the
proposal of the  spin-field-effect-transistor (SFET)  by Datta and Das more
than sixteen years ago.\cite{DD} In such a device,  ferromagnetic
material is used as source and drain for the injection and the
detection of spin-polarized electrons. The electron spins, moving
through a quasi-one-dimensional channel, precess about an
effective magnetic field tuned by a gate voltage. In $n$-type
zinc-blende semiconductors the local magnetic field gives origin
to a Zeeman-like splitting which, combined with the scattering,
can cause a spin relaxation/dephasing known as the
D'yakonov-Perel' (DP) mechanism.\cite{DP} In GaAs, the Dresselhaus
term\cite{gene} is dominant in the DP wave vector dependent
magnetic field, giving rise to a spin splitting due to bulk inversion
asymmetry. Furthermore in low-dimensional semiconductor structures
with asymmetric confining potential, another term
known as the Rashba term\cite{Rashba}
 contributes to the DP mechanism,
 giving rise to a spin
splitting based on the structure inversion asymmetry. While the
Rashba term is linear, the  Dresselhaus term is cubic in the wave
vector components.

A few years ago Schliemann {\em et al.} \cite{John} proposed a
non-ballistic spin-field-effect transistor based on the
competition of the Dresselhaus and the Rashba terms. In such a transistor
a gate voltage is tuned to give equal strengths to both terms,
leading to a very long spin dephasing time for the spin
polarization along (110) direction.\cite{John} In their work the
cubic term is argued to be unimportant.
Cheng and Wu studied the effect of the cubic term on the spin relaxation time
of spins along the (110) direction
by solving the kinetic spin Bloch equations, and obtained a finite spin
relaxation time.\cite{chengwu}
Actually, in spin transport,
 when spins are injected into a sample, in addition to the
direction of spin polarization, the direction of spin
diffusion/injection is also important. The spin relaxation
in transport depends on both directions. In the
present work,  the
anisotropy of the spin-orbit coupling is taken into account when
we study the ideal balance between the Rashba and the Dresselhaus terms in
a quasi-two-dimensional channel of a GaAs quantum well. Therefore,
both the direction of the spin polarization and the direction of
the spin injection are analyzed. We further
show that the cubic term appearing in the
Dresselhaus effective magnetic field cannot be neglected in the
competition with the Rashba field.
Our treatment goes beyond the  single
particle method in that we use a fully microscopic many-body approach
by solving the kinetic spin Bloch equations (KSBE).\cite{wu} As shown by
Weng and Wu, the correlations between the spin-up and spin-down
states, {\em i.e}., the off-diagonal terms of the density matrix
in spin space, play an essential role in the spin
diffusion/transport.\cite{wengwu1} Studying the spin polarizations
and the oscillations in the direction of diffusion one can
calculate the injection length as a function of the direction of
injection.

This paper is organized as follows: In Sec.\ \ref{so} we describe
the Rashba and the Dresselhaus spin-orbit couplings and study the
effective magnetic field under different injection directions in
the quantum well. In Sec.\ \ref{ke} we construct the KSBE
and we discuss how they change with the injection
direction. In Sec.\ \ref{res} we apply the KSBE to our
model and present the results for different parameters.
We conclude in Sec.\ \ref{fc}.

\begin{figure}
  \begin{center}\includegraphics[width=5.5cm]{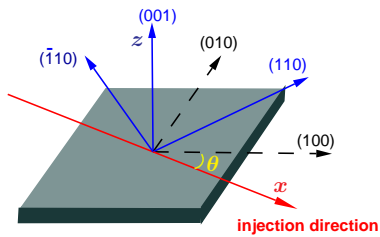}\end{center}
%\centerline{\psfig{figure=figciw1.eps,height=5.5cm,angle=0}}
\caption{(Color online) Schematic of the different directions considered
for the spin polarizations [(110), ($\bar{1}10$) and (001)-axes]
and spin diffusion/injection
($x$-axis).}
\label{schem}
\end{figure}

\section{Spin-orbit coupling}
\label{so}

The DP mechanism introduces a spin-orbit term in the Hamiltonian
which can be expressed by a Zeeman-like term with a
wave-vector-dependent effective magnetic field:
\begin{equation}
H_{so}=\mathbf \Omega(\mathbf
k)\cdot\mbox{\boldmath$\sigma$\unboldmath}.
\end{equation}
For convenience, the effective $g$-factor and  the Bohr magneton
are absorbed into the definition of $\mathbf \Omega(\mathbf k)$.
For GaAs, which lacks an inversion symmetry center, the effective
magnetic field contains the Dresselhaus term. For electrons in
infinite-barrier-height quantum wells with a small well width $a$ in the
$\Gamma_6$ band, this term is:\cite{winkler}
\begin{eqnarray}
{\bf \Omega}_{D}(\mathbf k)=\gamma \left(
\begin{array}{c}
k_x(k_y^2-\langle k_z^2\rangle)\\
k_y(\langle k_z^2\rangle-k_x^2)\\
0
\end{array}
\right), \label{dressel}
\end{eqnarray}
where $\gamma$ represents the spin splitting parameter and
  $\langle k_z^2 \rangle=(\pi/a)^2$.

If a space gradient is applied to the sample, another effect shows
up, the Rashba effect. The Rashba term in the Hamiltonian for
electrons in the $\Gamma_6$ band is
\begin{equation}
H_R=r^{6c6c}_{41}\mbox{\boldmath$\sigma$\unboldmath}\cdot\mathbf
k\times \mbox{\boldmath$\varepsilon$\unboldmath},
\end{equation}
where $ r^{6c6c}_{41}$ is a material parameter, and
$\mbox{\boldmath$\varepsilon$\unboldmath}$ is the electric field
determined by the asymmetry.\cite{winkler}

Taking the $z$-axis in the (001) direction ($z$ will be considered
throughout this paper as the growth direction of the quantum well)
and $\mbox{\boldmath$\varepsilon$\unboldmath}=\varepsilon
 \mathbf{\hat z}$, the contribution of the Rashba term to the
effective magnetic field may be written as
\begin{eqnarray}
{\bf \Omega}_{R}(\mathbf k)=\alpha
\left(
\begin{array}{c}
k_y\\
-k_x\\
0
\end{array}
\right),\label{rash}
\end{eqnarray}
where $\alpha$ is the Rashba parameter. Next, we consider the
influence on the effective magnetic field when we change the
injection direction  to $(\cos\theta,\sin\theta,0)$ and choose this
direction as the $x$-axis (A schematic of the configuration is
shown in Fig.\ \ref{schem}). The effective magnetic field changes
according to ${\bf\Omega}^{\prime}(\mathbf k)=U^{-1}{\bf \Omega}
(U\mathbf k)$, with
\begin{equation}
U=\begin{pmatrix}\cos\theta&\sin\theta&0\\
  -\sin\theta&\cos\theta&0\\ 0&0&1\end{pmatrix}.
\end{equation}
Therefore, we obtain for the Dresselhaus field:
\begin{widetext}
\begin{equation}
{\bf \Omega}^{\prime}_D(\mathbf k)=\gamma \langle
k_z^2\rangle\begin{pmatrix}-k_x\cos
2\theta +k_y\sin 2\theta\\
k_x\sin 2\theta + k_y\cos 2\theta\\
0\end{pmatrix} +\gamma(\frac{k_x^2-k_y^2}{2}\sin
2\theta+k_xk_y\cos 2\theta)
\begin{pmatrix} k_y\\
-k_x\\
0\end{pmatrix}\  ,
\label{dress}
\end{equation}
\end{widetext}
where the wave vector components are written in the rotated
coordinate system. For the Rashba field we obtain:
\begin{equation}
{\bf \Omega}^{\prime}_R(\mathbf k)=\alpha\begin{pmatrix}k_y\\ -k_x\\
0\end{pmatrix}\ .
\end{equation}
Notice that the Rashba field is expressed in the new coordinates
in the same way as in Eq.\ (\ref{dress}).

In the following, we discuss the diffusion in two different ways:
(i) We take the identical strengths of $\alpha=\beta$ with
  $\beta\equiv\gamma\langle k_z^2\rangle$ to study the diffusion direction
  dependence and the spin polarization dependence of the diffusion. In
  this case, the spin orbit coupling ${\bf \Omega}^\prime={\bf \Omega}_D^\prime
  +{\bf \Omega}_R^\prime$ is written as
\begin{widetext}
\begin{equation}
{\bf \Omega}^{\prime}(\mathbf k)=2 \beta \left(\sin(\theta-\frac{\pi}{4})k_x
+\cos(\theta-\frac{\pi}{4})k_y\right) \hat{\mathbf n}_0 + \gamma(\frac{k_x^2-k_y^2}{2}\sin
2\theta+k_xk_y\cos 2\theta)
\begin{pmatrix} k_y\\
-k_x\\
0\end{pmatrix}
\label{eq:total}
\end{equation}
\end{widetext}
with the special direction $\hat{\mathbf
  n}_0=\begin{pmatrix}\cos(\pi/4-\theta)\\ \sin(\pi/4-\theta) \\
  0\end{pmatrix}$ representing the crystal direction
$(110)$ in the rotated coordinates.
(ii) We study the effect of the gate voltage on the spin
diffusion.  We treat this issue in Sec.\ IV C.

\section{KSBE and Inhomogeneous Broadening}
\label{ke}

Our fully microscopic treatment concerns the calculation of the spin
density matrix for the electron with momentum $\mathbf k$ at
position $\mathbf r=(x,y)$:
\begin{eqnarray}
\rho_{\mathbf k}(\mathbf r,t)=
\begin{pmatrix}
f_{\mathbf k\uparrow} & \rho_{\mathbf k\uparrow\downarrow}\\
\rho_{\mathbf k\downarrow\uparrow} & f_{\mathbf k\downarrow}
\end{pmatrix}\ .
\end{eqnarray}
The diagonal elements $f_{\mathbf k\sigma}$ stand for the electron
distribution functions of spin $\sigma$ whereas the off-diagonal
elements  $\rho_{\mathbf k\uparrow\downarrow}=\rho^\ast_{\mathbf
k\downarrow\uparrow}$ represent the correlations
between the spin-up and -down states.

By using the non-equilibrium Green function method with gradient
expansion as well as the generalized Kadanoff-Baym ansatz,\cite{haug} we
construct the KSBE as follows,\cite{wengwu1} assuming the
diffusion to take place in the $x$-direction:
\begin{widetext}
\begin{equation}
\label{eq:BEQ}
\hbar\frac{\partial \rho_{\mathbf{k}}(x,
t)}{\partial  t} + e\frac{\partial \Psi(x, t)}{\partial
x}\frac{\partial \rho_{\mathbf k}(x, t)}{\partial k_x}
 + \frac{\hbar^2k_x}{m^*}\frac{\partial \rho_{\mathbf k}(x,t)}{\partial x}
+i[(g\mu_B\mathbf{B}+\mathbf{\Omega^{\prime}}(\mathbf
k))\cdot\frac{\mbox{\boldmath$\sigma$\unboldmath}
}{2}+{\cal E}_{\mathtt{HF}}(x,t), \rho_{\mathbf k}(x, t)] =\hbar
\left.\frac{\partial\rho_{\bf k}(x,t)}{\partial t}\right|_{\mathtt{s}}\ .
\end{equation}
\end{widetext}
Here $\Psi( x,t)$  is the electric potential
satisfying the Poisson equation
\begin{equation}
\nabla_{\mathbf r}^2\Psi(\mathbf r)=e[n(\mathbf r)- N_0(\mathbf
r)]/(a\kappa_0)\ ,
\label{possion}
\end{equation}
with $n(\mathbf r)$ standing for the electron density at position
$\mathbf r$, $N_0(\mathbf r)$ representing the background positive
charge density and $\kappa_0$ being the static dielectric function.
${\cal E}_{\mathtt{HF}}(x,t)$ is the Hartree-Fock term from the
Coulomb interaction.
The scattering terms $\left.\frac{\partial\rho_{\bf k}(x,t)}
{\partial t}\right|_{\mathtt{s}}$ include
the electron-electron and electron-phonon  scattering. Their expressions can be
found in Ref.\ \onlinecite{wengwu2}.
Details of the numerical scheme as well as the material parameters
 are laid out in Ref.\ \onlinecite{newchengwu}. It has been shown
 that Coulomb scattering plays a very important rule in spin
dephasing\cite{wengwu2,chengwu,prbclv,zhou} and spin
diffusion/transport\cite{newchengwu} and therefore cannot be neglected in
calculating the spin diffusion length.

When the system reaches its steady state, in the absence of an
external  field, it can be described by the equation
\begin{widetext}
\begin{eqnarray}
\label{eq:BEQ1} \frac{\hbar^2k_x}{m^{\ast}}\frac{\partial
\rho_{\mathbf k}(x,t)}{\partial x}
+i[\mathbf{\Omega^{\prime}}(\mathbf
k)\cdot\frac{\mbox{\boldmath$\sigma$\unboldmath}
}{2}+{\cal E}_{\mathtt{HF}}(x,t), \rho_{\mathbf k}(x, t)] =
\left.\hbar\frac{\partial\rho_{\mathbf k}(x,t)}{\partial
    t}\right|_{\mathtt{s}}\ .
\end{eqnarray}
\end{widetext}
By neglecting the Hartree-Fock and the scattering terms,
Eq.\ (\ref{eq:BEQ1}) can be solved analytically.\cite{newchengwu} After
dividing both sides of the equation  by $k_x$, we see that along with the diffusion, the
spin polarization for each wave vector $\mathbf k$ precesses along
$\mbox{\boldmath$\omega$\unboldmath}_{\mathbf
k}=\frac{m^{\ast}}{2\hbar^2k_x}{\bf \Omega}^{\prime}({\bf k})$. The fact
that spins with different momentums precess with different
frequencies is referred to as inhomogeneous broadening.\cite{wuning,alleneber}
With any spin-conserving
scattering included, the inhomogeneous broadening results in spin
dephasing.\cite{wuning,wengwu1}

For $\alpha=\beta$, $\mbox{\boldmath$\omega$\unboldmath}_{\mathbf k}$ reads
\begin{widetext}
\begin{equation}
\mbox{\boldmath$\omega$\unboldmath}_{\mathbf k}=\frac{m^{\ast}}{2\hbar^2} \left\{2\beta
\left(\sin(\theta-\frac{\pi}{4})+\cos(\theta-\frac{\pi}{4})\frac{k_y}{k_x}\right)
\hat{\mathbf n}_0 + \gamma(\frac{k_x^2-k_y^2}{2}\sin
2\theta+k_xk_y\cos 2\theta)
\begin{pmatrix} k_y/k_x\\
-1\\
0\end{pmatrix}\right\}\ .
\label{omega}
\end{equation}
\end{widetext}
We find that $\mbox{\boldmath$\omega$\unboldmath}_{\mathbf k}$ includes two parts: the zeroth-order
term (on $k$) which is always along the same direction $\hat{\mathbf
  n}_0$, and the second-order term.  The zeroth-order term includes
both the $\mathbf k$-dependent and the $\mathbf k$-independent terms, while
the second-order term is always $\mathbf k$-dependent. According to
the previous works,\cite{wuning,wengwu1,wengwushi,newchengwu} the $\mathbf
k$-dependent term contributes to the inhomogeneous broadening.

We first analyze the spin diffusion without the
third-order term of the Dresselhaus term (and
hence the second-order term of $\mbox{\boldmath$\omega$\unboldmath}_{\bf k}$).
In this case, the effective magnetic field for
each wave vector $\mathbf k$  points to the same direction
  $\hat{\mathbf n}_0$. From the previous works,\cite{John,chengwu} we know that
the  spin dephasing in the time domain, in the spatial homogeneous case,
shows strong anisotropy with respect to
the direction of spin polarization: The spin
polarization along the direction $\hat{\mathbf n}_0$  has an
infinite dephasing time, while the ones vertical to this direction
have a very short spin dephasing time due to the large
spin-orbit coupling. This is because
when the spin polarization direction is the same as the direction of the
effective  magnetic field, the spin polarization cannot precess and
hence there is no
inhomogeneous broadening. Consequently  the spin polarization has an infinite
dephasing time. The same is true for the spin diffusion:
When the spin polarization is along the same direction of
$\mbox{\boldmath$\omega$\unboldmath}_{\mathbf k}$, {\em i.e.},
$\hat{\mathbf n}_0$ (the crystal direction $(110)$),
the spin polarization will not decay {\em regardless of the direction of
the spin injection}, even in the presence of
scattering, as there is no spin precession and hence no inhomogeneous
broadening; However, when the spin polarization
is perpendicular to $\hat{\mathbf n}_0$,
the spin diffusion length is very short due to the large
inhomogeneous broadening.

Nevertheless, in contrast to the spin dephasing in a spatially homogeneous
system, the spin diffusion has an extra degree of freedom,
{\em i.e.}, the spin diffusion/injection
direction (which should not be confused with the spin polarization
direction). This degree of freedom introduces an additional level of anisotropy
to the spin diffusion/transport.
When the spin diffusion/injection direction is along $\theta=3\pi/4$,
{\em i.e.}, ($\bar110$) direction,
the precession frequency
$\mbox{\boldmath$\omega$\unboldmath}_{\mathbf k}=\frac{m^{\ast}\beta}{\hbar^2}
  \hat{\mathbf n}_0$ becomes $\mathbf
  k$-independent if the cubic term of the Dresselhaus term is neglected.
Moreover, even with the scattering included,
Eq.\ (\ref{eq:BEQ1}) can also be satisfied by the following solution:
\begin{equation}
\label{eq:BEQ2}
\rho_{\mathbf k}(x) =e^{im^{\ast}\gamma\langle
 k_z^2\rangle\sigma_y x/\hbar^2} \rho_{\mathbf
 k}(x=0)e^{-im^{\ast}\gamma\langle k_z^2\rangle\sigma_y x/\hbar^2}
\end{equation}
where $\rho_{\mathbf k}(x=0)$ is the Fermi distribution with the
spin polarization at the left boundary.
We see from Eq.\ (\ref{eq:BEQ2}) that in this special
case the oscillation period
is the same for all $\mathbf k$'s, with the value
 $\frac{\pi\hbar^2}{m^{\ast}\gamma \langle k_z^2\rangle}$. So,
  there is no inhomogeneous broadening and the spin has an infinite
  diffusion length, {\em regardless of the
  direction of the initial spin polarization}. This result is mainly
induced by the peculiar precession frequency
$\mbox{\boldmath$\omega$\unboldmath}
_{\mathbf k}$ which is determined by the  spin-orbit coupling and the
diffusion velocity $k_x$, so it cannot be obtained in the
spatially homogeneous case where the precession frequency is only
 determined by the spin-orbit coupling. Of course, when
we include the cubic term
 in the spin-orbit coupling, the oscillation
 period becomes wave-vector dependent. This inhomogeneous
 broadening, then, results in a finite diffusion length.

In brief, we stress the importance of the
inhomogeneous broadening caused by the diffusion term as compared to
the spin dephasing in the spatially homogeneous case.
Consequently the spin
polarization shows strong anisotropy not
only for the direction of the initial spin polarization but also for
the direction of the spin
diffusion. Moreover, the cubic Dresselhaus term cannot be neglected.
In what follows we solve the problem by using the
fully microscopic many-body approach of the KSBE with the cubic
Dresselhaus term and the scattering explicitly included.

\begin{figure}
  \begin{center}\includegraphics[height=5.5cm]{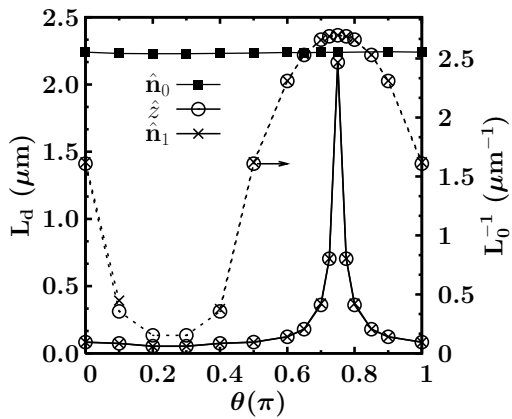}\end{center}
%\centerline{\psfig{figure=figciw1.eps,height=5.5cm,angle=0}}
\caption{Spin diffusion length $L_d$ (solid curves) and the inverse of
the spin oscillation period $L_0^{-1}$ (dashed curves)
for $\alpha=\beta$ as functions of the injection
direction for different spin polarization directions
$\hat{\mathbf n}_0$, $\hat{\bf z}$ and $\hat{\mathbf n}_1$ at
$T=200$\ K. It is noted that the scale of the spin oscillation
period is on the right hand side of the frame.}
\label{fig1}
\end{figure}

\section{Numerical Results}
\label{res}

The KSBE are  solved numerically  for (001) GaAs quantum wells of width
$a=5$\ nm including the electron-electron and
the electron--LO-phonon scattering.  The value of the Dresselhaus
coefficient is  as $\gamma =
25$\  eV$\cdot$\AA$^3$.\cite{jusse} The electron density and the
temperature are taken as $N_e=4.0\times10^{11}$\ cm$^{-2}$ and
$T=200$\ K separately unless otherwise specified.  All matrix
elements of the interactions are given in Ref.\ \onlinecite{wengwu2}.
For the spin polarization along direction $\hat{\mathbf{n}}$,
 the boundary conditions are given by\cite{chengwu}
\begin{equation}
\begin{cases}
\left.\rho_{\mathbf k}(x=0,t)\right|_{k_x>0}
= \frac{F_{\mathbf k,\uparrow}+F_{\mathbf k,
\downarrow}}{2}+\frac{F_{\mathbf k,\uparrow}-F_{\mathbf k,
\downarrow}}{2}\hat{\mathbf
n}\cdot\mbox{\boldmath$\sigma$\unboldmath}\ ,\\
\left.\rho_{\mathbf k}(x=L, t)\right|_{k_x<0} = \frac{F_{\mathbf k,\uparrow}+F_{\mathbf k,
    \downarrow}}{2}\ ,
\end{cases}
\end{equation}
with the Fermi distribution $F_{\mathbf
  k,\sigma}=[e^{(k^2/2m^{\ast}-\mu_{\sigma})/k_BT}+1]^{-1}$
($\sigma=\uparrow\downarrow$) and the
chemical potential $\mu_\sigma$ determined by the polarized electron
density. The diffusion length and the oscillation period are extracted
from the spatial evolution of the spin polarization along $\hat{\mathbf n}$
direction $\Delta N(x)=\sum_{\mathbf k}\mbox{Tr}[\rho_{\mathbf
  k}(x)\hat{\mathbf n}\cdot\mbox{\boldmath$\sigma$\unboldmath}]$.

\subsection{Spin-diffusion/injection--direction and
spin-polarization--direction dependence at $\alpha=\beta$}

We first fix the spin polarization along the (110) direction
($\hat{\mathbf n}_0$) and
study the spin diffusion length as a function of the spin injection
direction in the presence of the
cubic Dresselhaus term. For comparison, we also study the
cases when the spin polarizations
are perpendicular to the (110) direction, {\em i.e.}, $\hat{\bf z}$ and
$\hat{\mathbf n}_1=\hat{\bf z}\times\hat{\mathbf n}_0$ and show
how the spin injection lengths change as a function of spin diffusion/injection
direction. Subsequently we explore the special case
when the spin diffusion/injection
direction is along ($\bar110$), {\em i.e.}, $\theta=3\pi/4$ and show
how the spin injection length changes with the spin polarization
in the presence of the cubic Dresselhaus term.
The spin polarization calculated from the KSBE Eq.\ (\ref{eq:BEQ})
and the Poisson Eq.\ (\ref{possion}) along these
three spin polarizations can be well fitted by
\begin{equation}
\Delta N(x)=C\exp(-\frac{x}{L_d})\cos(\frac{2\pi
x}{L_0}+\phi)\ ,
\label{eq:fitexpos}
\end{equation}
where $L_d$ is the spin diffusion length and $L_0$ represents the
oscillation period.

In Fig.\ \ref{fig1} the spin diffusion length $L_d$ and the inverse of the
 spin oscillation period $L_0^{-1}$ are plotted against the
spin diffusion/injection angle $\theta$ for spin polarizations
along $\hat{\bf n}_0$,  $\hat{\bf z}$ and
$\hat{\mathbf n}_1$, respectively. It is interesting to see that
with the inclusion of the cubic Dresselhaus
term, the spin injection length becomes finite but still {\em independent}
on the direction of the spin diffusion/injection if the  spin polarization
is along (110) ($=\hat{\bf n}_0$). This is in agreement with the
case without the cubic term where $L_d$ becomes infinite.
At the same time, the spin polarization along this direction has an
infinite oscillation period $L_0=\infty$. This can be understood
because the spin polarization is in the same direction as the effective
magnetic field given by the zeroth-order term in
$\mbox{\boldmath$\omega$\unboldmath}_{\mathbf k}$
and, thus, results in no oscillations. It is noted that
the effective magnetic field from the second-order term
in $\mbox{\boldmath$\omega$\unboldmath}_{\mathbf k}$ is ${\bf k}$-dependent
and therefore cannot lead to any oscillation at high temperature due to the
scattering.\cite{chengwu,harley,prbclv} However, this second-order term causes an inhomogeneous
broadening which leads to a finite spin diffusion length.
By rewriting the second-order term of
$\mbox{\boldmath$\omega$\unboldmath}_{\mathbf k}$ into
$\frac{m^{\ast}\gamma}{2\hbar^2}\frac{k^2}{2}
\sin(2\theta_{\mathbf k}+2\theta)\begin{pmatrix}ky/kx \\
-1 \\ 0\end{pmatrix}$ with $k$ and $\theta_{\mathbf k}$ denoting the
magnitude and the direction of the wave vector $\mathbf k$ separately, one
finds that the magnitude of the inhomogeneous broadening does not change with
the spin diffusion/injection direction $\theta$. Consequently
the spin diffusion length does not change with the spin
diffusion/injection direction.

Strong anisotropy is again found
for spin polarizations along the  directions perpendicular to
$\hat{\bf n}_0$ except for the special
case when $\theta=3\pi/4$. Moreover, it is found in the figure that,
for the perpendicular directions, both $L_d$ and $L_0$ depend sensitively
on the spin diffusion/injection direction (additional anisotropy).
It is further noticed that $L_d$ and $L_0$ are almost identical for both
the perpendicular directions $\hat {\bf z}$ and $\hat{\bf n}_1$.
These results can be understood again from the inhomogeneous
broadening induced by $\mbox{\boldmath$\omega$\unboldmath}_{\mathbf k}$.
As $\gamma$ in Eq.\ (\ref{omega}) is much smaller than $\beta$, the
inhomogeneous broadening for spin polarizations  $\hat {\bf z}$
and $\hat{\bf n}_1$ are therefore defined by the zero-th order term
of $\mbox{\boldmath$\omega$\unboldmath}_{\mathbf k}$, which is
the same for both perpendicular directions and is spin
diffusion/injection direction $\theta$ sensitive.
The oscillation period in the diffusion $L_0$ is determined by the
$\mathbf k$-independent component of
$\mbox{\boldmath$\omega$\unboldmath}_{\mathbf k}$,\cite{newchengwu} {\em i.e.},
${m^{\ast}\beta\sin(\theta-\pi/4)}/{\hbar^2}$ which is in good agreement
with the results in  Fig. \ref{fig1}.

Now we turn to the case with special fixed spin diffusion/injection
direction $\theta=3\pi/4$, and investigate the spin
diffusion length with different spin polarizations. It is interesting to
see from Fig.\ \ref{fig1} that similar to the case without the cubic
Dresselhaus term, the spin diffusion lengths for the three
spin polarizations are almost identical. Therefore the
anisotropy disappears for this particular spin diffusion/injection direction.
This is because when the spin diffusion/injection direction
is $\theta=3\pi/4$, the  $\mathbf k$-dependent component
in the zero-th order term of $\mbox{\boldmath$\omega$\unboldmath}_{\mathbf k}$
disappears. This results in the same inhomogeneous broadening
for any spin polarization.

\subsection{Temperature and electron-density dependence with
injection direction along  $(\bar110)$ and $\alpha=\beta$}

\begin{figure}
  \begin{center}\includegraphics[height=5.5cm]{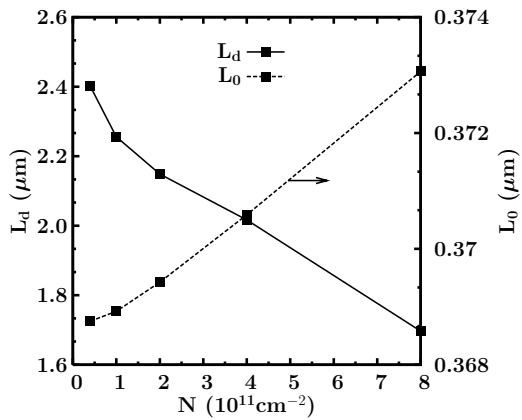}\end{center}
%  \centerline{
%  \psfig{figure=figciw2.eps,height=5.5cm,angle=0}}
\caption{Spin diffusion length $L_d$  and spin oscillation period $L_0$
 at $\theta=3\pi/4$ and $T=200$\ K as functions of the
  electron density. Note the scale of the oscillation period is on the
right hand side of the frame.}
  \label{fig2}
\end{figure}

\begin{figure}
  \begin{center}\includegraphics[height=5.5cm]{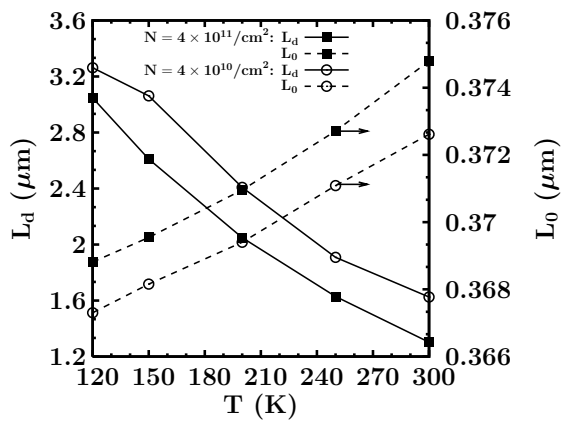}\end{center}
%  \centerline{
%  \psfig{figure=figciw3.eps,height=5.5cm,angle=0}}
\caption{Spin diffusive length $L_d$  and spin oscillation period $L_0$
at two different electron densities,
$4\times 10^{11}$\ cm$^{-2}$ and
 $4\times 10^{10}$\ cm$^{-2}$ {\em vs.} the temperature. The spin
diffusion/injection direction $\theta=3\pi/4$. Note the scale of the
oscillation period is on the right hand side of the frame.}
\label{fig3}
\end{figure}

Now we investigate the temperature and electron density dependence of the
spin diffusion length at the special spin diffusion/injection direction
$\theta=3\pi/4$ with $\alpha=\beta$. As shown in the previous
section the spin diffusion is isotropic with respect to the spin polarization
direction. Hence we choose the
spin polarization to be along the  $z$-axis. In Fig.\ \ref{fig2}
the oscillation period and the diffusion length are plotted as
functions of the electron density. One finds that
the diffusion length $L_d$ decreases with the density, whereas the
oscillation period $L_0$ increases with it. However, it is noted
that  the change observed in the oscillation period
is almost negligible (1\ \%) in contrast to the
pronounced changes observed in the diffusion length (of the order of 40\ \%).
We further investigate the temperature dependence
of the spin diffusion length and the oscillation period in
Fig.\ \ref{fig3} by taking two typical electron densities, {\em i.e.},
$N=4\times 10^{10}$\ cm$^{-2}$ and $4\times 10^{11}$\ cm$^{-2}$.
Similar behavior is found here compared to the density dependence:
the spin diffusion length $L_d$ decreases markedly with the
temperature whereas the oscillation period $L_0$ only increases
slightly with it.

Both dependences above can be again understood  by the
inhomogeneous broadening.
Unlike the results in Ref.\ [\onlinecite{newchengwu}] where only the
Dresselhaus term is considered and a very weak
temperature dependence is obtained, the diffusion length here has a
strong  temperature dependence. This can
be understood by the difference in the behavior of the inhomogeneous
broadening, which is determined by the anisotropy of the
precession frequency  $\mbox{\boldmath$\omega$\unboldmath}_{\mathbf k}$.
In the previous case, the zero-th order term dominates the
inhomogeneous broadening which changes little with
temperature and electron density. However, in the
present case with $\alpha=\beta$ and $\theta=3\pi/4$, the
$\mathbf k$-dependent zeroth-order term is always zero
and the second-order term alone
determines the inhomogeneous broadening. This term increases
effectively with the
temperature and with the electron density due to the increase of the
average value of $k^2$. This is the reason for the obtained marked decrease
of the spin diffusion length.  The oscillation period $L_0$
is determined by the $\mathbf k$-independent zeroth-order term of
$\mbox{\boldmath$\omega$\unboldmath}_{\mathbf k}$,
 which is independent of the
electron density and the temperature. Therefore one
observes only a  slight change in the
oscillation period, originating from
the second-order term and the scattering.

\begin{figure}
  \begin{center}\includegraphics[height=5.5cm]{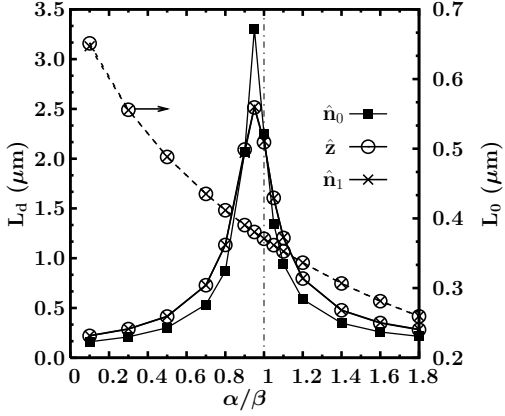}\end{center}
%  \centerline{
%    \psfig{figure=figciw4.eps,height=5.5cm,angle=0}}
\caption{Spin diffusion length $L_d$ (solid curves)
 and spin oscillation period $L_0$ (dashed curves)
as functions of the ratio $\alpha/\beta$ with the
injection direction $\theta=3\pi/4$ for different spin
polarization directions  $\hat{\mathbf n}_0$, $\hat{\bf z}$ and
$\hat{\mathbf n}_1$ at $T=200$\ K. The chain line is used to guide the eyes.
The scale of the oscillation period is on the right hand side of the frame.}
\label{fig4}
\end{figure}

\subsection{Gate-voltage dependence with injection direction
along  $(\bar110)$}

We now tune the gate voltage in order to change the relative
importance of the Dresselhaus and the Rashba terms. The spin
diffusion/injection direction is again fixed at  $\theta=3\pi/4$.
The calculated spin diffusion length $L_d$ and the spin
oscillation length $L_0$ are plotted against $\alpha/\beta$
in Fig.\ \ref{fig4} at $T=200$\ K, with the spin polarizations being along
$\hat{\bf n}_0$, $\hat{\bf z}$ and $\hat{\mathbf n}_1$ respectively.
It is noted that since $L_0=\infty$ when the spin polarization
is along  $\hat{\bf n}_0$ [(110)], we only plot $L_0$ for the
two perpendicular directions in the figure.

Three main results are obtained: (i) Differing from the
case without the cubic Dresselhaus term, the
maximum diffusion length occurs interestingly at $\alpha/\beta\sim0.95$
(instead of 1) for each spin polarization direction;
(ii) The isotropy of the spin polarization direction at
$\alpha/\beta\neq 1$ is suppressed except at $\alpha/\beta\sim
0.9$, and the diffusion lengths for the spin
polarization along $\hat{\bf z}$ and $\hat{\bf n}_1$ are always
identical; (iii) The spin oscillation period $L_0$ decreases with $\alpha/
\beta$. These results can also be understood from the inhomogeneous
broadening. For $\theta=3\pi/4$, the spin precession frequency
along the diffusion direction is given by
\begin{eqnarray}
\mbox{\boldmath$\omega$\unboldmath}_{\mathbf k}
 &=& [(-\beta+\alpha +\frac{\gamma k^2}{2}
)\frac{k_y}{k_x} - \gamma k_xk_y]\hat{\mathbf n}_1 \nonumber
\\
&&\mbox{} - [\beta+\alpha -
\frac{\gamma}{2}(k_x^2-k_y^2)]\hat{\mathbf n}_0\ .
\label{3piov4}
\end{eqnarray}
At $T=200$\ K, $\langle k^2\rangle/2\sim 0.11 (\pi/a)^2$ with
$\langle\cdot\rangle$ representing the average over the imbalance of
the spin-up and -down electrons. Therefore,
$(-\beta+\alpha +\frac{\gamma
k^2}{2})\frac{k_y}{k_x}\hat{\mathbf n}_1$ is almost zero at
$\alpha/\beta\sim0.95$ and does not
contribute to the inhomogeneous broadening. This
results in a maximum of the spin diffusion length.
It is further noted that  the inhomogeneous broadening
term $\frac{\gamma}{2}(k_x^2-k_y^2)\hat{\mathbf
 n}_0$ gives a larger effect when the spin polarization is along $\hat{\bf z}$
and $\hat{\mathbf n}_1$ than that when the
spin polarization is along $\hat{\mathbf n}_0$ as the latter coincides with
the precession axis.  This explains why one obtains
a larger diffusion length when the spin polarization is along
$\hat{\bf n}_0$ at $\alpha/\beta\sim0.95$.
Finally as the spin oscillation period $L_0$ is determined by the magnitude of
the ${\bf k}$-independent term of
$\mbox{\boldmath$\omega$\unboldmath}_{\mathbf k}$, {\em i.e.}, $\beta
(\alpha/\beta+1)\hat{\bf n}_0$, which increases with $\alpha/\beta$.
Therefore, $L_0$ decreases with $\alpha/\beta$.

\section{Conclusions}
\label{fc}

In conclusion, we study the spin diffusion in (001) GaAs quantum wells
with competing Dresselhaus and
Rashba spin-orbit coupling strengthes by solving the KSBE
with the electron-phonon and
the electron-electron Coulomb scattering explicitly included.
It is shown that unlike the spin dephasing in the time domain where strong
anisotropy is determined by different spin polarization, for spin diffusion
it is also determined by the spin diffusion/injection direction.
By neglecting the cubic term of the Dresselhaus spin-orbit coupling and
with $\alpha=\beta$, the ideal case of an
infinite diffusion length is obtained {\em either}
for the spin polarization along
$(110)$ regardless of the spin diffusion direction {\em or} for the spin
diffusion direction along $(\bar{1}10)$ regardless of the spin
polarization direction.
However, the cubic term cannot be neglected,
resulting in a finite diffusion length, which in fact is small,
about 2\ $\mu$m at the temperature of 200\ K when the electron
density lies between $4\times 10^{10}$\ cm$^{-2}$ and
$4\times 10^{11}$\ cm$^{-2}$.
It is then shown that when $\alpha=\beta$, for the spin polarization along
$(110)$, the spin diffusion length changes little for the spin
diffusion direction, whereas for the other two perpendicular spin polarization
directions $(\bar{1}10)$ and $(001)$, the spin diffusion length shows
strong anisotropy for the spin diffusion/injection direction and has
 a peak when the spin
diffusion/injection direction is along $(\bar110)$.
When the spin diffusion/injection direction is along $(\bar110)$, the
 spin diffusion
length is isotropic for spin polarization. The electron density and temperature
dependence of the spin diffusion is closely investigated.

We also tune the gate voltage to show the competing effect
of the Dresselhaus and
the Rashba strengthes. It is found that with the cubic
 Dresselhaus term included,
the maximum spin diffusion length appears at $\alpha/\beta\sim0.95$
instead of 1
where the cubic term is neglected.
 The spin diffusion length decreases when the ratio of
the Rashba and Dresselhaus strengths deviates from 0.95.
These results can be well
understood by the inhomogeneous broadening.

\begin{acknowledgments}
This work was supported by the Natural Science Foundation of China
under Grant No.\ 10574120, the National Basic
Research Program of China under Grant No.\ 2006CB922205,
the Knowledge Innovation Project of Chinese Academy of
Sciences and SRFDP. One of the authors (I.C.C.L.) was partially
supported by an ESN from CNPq from Brazil. We would like to thank D.
Csontos for his critical reading of this manuscript.

\end{acknowledgments}

\end{document}